\documentstyle[prb,preprint,tighten,aps]{revtex}
\begin{document}
\draft
\title{Composite fermion edge states, 
fractional charge and current noise} 
\author{George Kirczenow\footnote{e-mail address: kirczeno@sfu.ca}}
\address{Department of Physics, Simon Fraser University, 
Burnaby, B.C., Canada  V5A 1S6}
\maketitle

\begin{abstract}
A composite fermion edge state theory of current 
fluctuations, fractional quasiparticle 
charge and Johnson-Nyquist noise in the fractional quantum
Hall regime is presented. It is shown that composite
fermion current fluctuations and the charges of the associated 
quasiparticles are strongly renormalized by 
the interactions between composite fermions. The important 
interaction is that mediated by the fictitious electric field 
associated with composite fermion currents. The dressed current 
fluctuations and quasiparticle charges are calculated 
self-consistently in a mean field theory for smooth edges. 
Analytic results are 
obtained. The values of the fractional quasiparticle 
charges obtained 
agree with the predictions of previous theories in the 
incompressible regions of the 2DEG
where those theories apply. In the compressible 
regions the magnitudes of the quasiparticle charges vary 
with position. 
Since Johnson-Nyquist noise arises 
from the compressible
regions, it is due to quasiparticles whose charges
differ from the simple fractions of $e$ that apply
in the incompressible regions. Never the less, 
the Nyquist 
noise formula $S=4{k_B}T G$ is obeyed on fractional
quantum Hall plateaus. Some implications for the interpretation of recent
shot noise measurements in the fractional quantum Hall 
regime are briefly discussed. 
\end{abstract}

\section{Introduction}
It was first shown by Laughlin,\cite{Laughlin} that the fractional 
quantum Hall effect occurs because electron-electron 
interactions result in incompressible states of two-dimensional 
electron systems at special filling fractions of a Landau level.
Subsequently, Jain showed 
that these incompressible many-body 
states can be understood, in a mean field sense, as arising 
from the spectral gaps between the single-particle Landau levels 
of quasi-particles known as composite fermions and that 
the fractional quantum Hall effect can be viewed as the integer 
quantum Hall effect of composite fermions.\cite{Jain} The composite 
fermion theory has yielded many remarkable results 
and there is now an extensive theoretical and 
experimental literature supporting it and examining its
ramifications.\cite{reviews} 

Following Jain's suggestion that the fractional quantum Hall 
effect is the integer 
quantum Hall effect of composite fermions,
a model of composite fermion edge 
states has been proposed\cite{edge}$^,$\cite{scat} 
that generalizes the very successful 
edge state theories of transport in the integer quantum Hall 
regime\cite{Halperin}$^-$\cite{Buttiker}
to fractional quantum Hall phenomena. 
This mean field model was developed for high mobility
systems with smooth edge potentials and has been able to
account for a great deal of experimental data
including the observed integer\cite{Kli} 
and fractional\cite{Tsui} quantized Hall conductances, 
the results of transport 
experiments on Hall bars with smooth potential barriers and 
constrictions,\cite{Ch2term}$^-$\cite{frost2} 
and the observed Fermi liquid-like behavior of 
Aharonov-Bohm resonances associated with antidots in the fractional 
quantum Hall regime.\cite{Franklin}$^,$\cite{Maasilta}

The singular gauge 
transformation\cite{Jain}$^,$\cite{LF}$^,$\cite{HLR} 
that transforms 
electrons into composite fermions by attaching to them tubes
of fictitious magnetic flux preserves charge, and therefore
the charge of a composite fermion is equal to that of an 
electron. However, it was pointed out by Goldhaber and Jain
\cite{GoJa} that the local charge associated with the composite
fermion is dressed by the presence of the other 
composite fermions in the 
system and thus becomes fractional, the fractions being
in agreement 
with the fractional quasiparticle 
charges predicted earlier by Laughlin and 
others.\cite{Laughlin}$^,$
\cite{Haldane}$^,$\cite{Halperinfrac} For example, for the fractional
filling $\nu=1/3$ of the lowest Landau level the dressed
quasiparticle charge $-e^*$ equals $-e/3$.

Recently experiments have been reported measuring current 
noise in the fractional quantum Hall regime\cite{Saminadayar}$^,$
\cite{de-Picciotto} as a probe of the fractional quasiparticle
charge. The results have been interpreted\cite{Saminadayar}$^,$
\cite{de-Picciotto} as direct 
experimental evidence of fractionally charged quasiparticles
with $e^*=e/3$ at $\nu=1/3$. However,
the theoretical predictions\cite{Laughlin}$^,$\cite{GoJa}$^-$
\cite{Halperinfrac} that $e^*=e/3$ at $\nu=1/3$ are based on the 
assumption that the electronic state is {\em incompressible},
whereas current noise originates at the edges of the 
sample where both incompressible and compressible regions
occur. Furthermore Johnson-Nyquist noise arises {\em entirely}
from the {\em compressible} regions. It is therefore
of interest to examine both the fractional quasiparticle 
charges and the associated current noise theoretically 
within an edge state model
that admits both incompressible and compressible regions 
at the edge. This is the purpose of the present article.

Section \ref{Model} contains a brief summary of
those aspects of 
composite fermion mean field theory and of the 
edge state model\cite{edge}$^,$\cite{scat}
that will be used in the remainder of this paper.

In Section \ref{Fluct} I discuss electric current 
fluctuations in 
the fractional quantum Hall regime within the 
framework of the 
edge state model,\cite{edge}$^,$\cite{scat} and show that
to understand these fluctuations one must consider the 
interactions between the composite fermions, the 
important interactions being those mediated 
by the fictitious electric field\cite{ZHK}$^,$\cite{RB}$^,$
\cite{Z}$^,$\cite{GoJa} 
associated with composite fermion currents. These
interactions renormalize the current fluctuations and 
their effects 
must be calculated self-consistently. This is done 
analytically in a
mean field approximation and it is
found that the dressing of the current fluctuations can be 
interpreted as a dressing of the charges of the 
composite fermion
quasi-particles. The dressed composite fermion 
charges are fractional and in incompressible regions
the predicted fractions 
agree with those obtained earlier 
by Laughlin and others.\cite{Laughlin}$^,$
\cite{Haldane}$^,$\cite{Halperinfrac}$^,$\cite{GoJa}. However 
in the compressible regions near the edge
the fractions are predicted to be different and 
to vary smoothly with position.
As the edge is approached and the local electron density tends 
to zero the dressed composite fermion charge 
approaches the electron
charge $-e$.

Current noise in mesoscopic systems has 
been discussed theoretically in
recent years by many authors,
\cite{Khlus}$^-$\cite{vO} 
but not from the perspective 
of composite fermion edge states. 
In Section \ref{Noise} I calculate the 
Johnson-Nyquist noise on the fractional quantum 
Hall plateaus using the composite 
fermion edge state model\cite{edge}$^,$\cite{scat} and the results
of Section \ref{Fluct} for the dressed current fluctuations. 
The method used
is a generalization of the wave packet arguments of Landauer and
Martin\cite{Land}$^,$\cite{LM} to 
composite fermion edge state theory and the fractional 
quantum Hall regime.
Summing the contributions of all of the 
active and silent modes\cite{edge}
of the edge state model and including the variation of the 
dressed quasiparticle charge across the edge 
obtained in Section \ref{Fluct} yields a result
in agreement with the Nyquist formula $S=4k_B T G_H$ for all of
Jain's filling fractions of a Landau level. Here $S$ is the
Johnson-Nyquist noise on the fractional quantum Hall plateau, 
$k_B$ is Boltzmann's constant, $T$ is the temperature and $G_H$
is the fractional quantized Hall conductance.

The significance of these results is discussed in 
Section \ref{Conc} where I also comment briefly on the 
implications for the interpretation of recent shot
noise experiments.

\section{Edge State Model}\label{Model}

In composite fermion theory, a singular gauge transformation attaches 
a tube of fictitious magnetic flux with an even number of flux 
quanta to each electron.\cite{Jain}$^,$\cite{LF}$^,$\cite{HLR}
The electrons together with the attached flux tubes obey Fermi 
statistics and are called ``composite fermions."\cite{Jain} In mean 
field theory the interactions between composite fermions that are 
due to the vector potentials associated with the tubes of fictitious 
flux are replaced by interactions with a fictitious average 
magnetic field and a fictitious electric field. 

The fictitious 
average magnetic field  is given by
\cite{Jain}$^,$\cite{LF}$^,$\cite{HLR}
\begin{equation}
{\bf B}_g = - n_e \hat{\bf B} m {h \over e} = -m \nu {\bf B}
\label{Bg}
\end{equation}
where an even integer number $m$ of 
flux quanta $h/e$ are attached to each electron, $n_e$ is
the two-dimensional electron density, $\hat{\bf{ B}}$ is the unit
vector pointing in the direction of the true magnetic field 
$\bf{ B}$ and 
$\nu = n_e h/eB$
is the Landau level filling parameter. Thus the
composite fermions experience an effective magnetic field
\begin{equation}
{\bf B}_{eff} = {\bf B} + {\bf B}_g
\label{Beff}
\end{equation}
and fill effective Landau levels spaced in energy by 
$\hbar \omega_{eff}$ where 
\begin{equation}
\omega_{eff}=e|{\bf B}_{eff}|/m^* 
\label{omega}
\end{equation}
is the effective
cyclotron frequency and $m^*$ is the composite fermion effective 
mass.
The fictitious electric field 
is\cite{ZHK}$^,$\cite{RB}$^,$\cite{Z}$^,$\cite{GoJa} 
\begin{equation}
{\bf E}_g = -({\bf J} \times \hat{\bf B})mh/e^2 
\label{Eg}
\end{equation}
where ${\bf J}$ is the two-dimensional electric current density.

In the composite fermion edge state model\cite{edge}$^,$\cite{scat} 
it is assumed that the electron density is slowly varying with 
position and that, for local Landau level filling fractions in 
the vicinity of $1/m$, the composite fermion Landau level energies 
behave qualitatively like 
\begin{equation}
E_{m,r}=\left( r+{1 \over 2}\right)\hbar\omega_{eff}+W
\label{cflands}
\end{equation}
where $r = 0, 1, 2,...$ and $W$ is 
the position-dependent 
composite fermion effective potential energy that includes
the effects of the fictitious electric field. 
Equation (\ref{cflands}) is a good description of the composite 
fermion Landau level structure in uniform systems\cite{Jain} 
and yields edge states that propagate in the direction 
consistent with experiments.\cite{edge}$^,$ \cite{Moon} 
This behavior of the composite fermion Landau 
levels near an edge is illustrated in Fig. \ref{f1}. 
The effective magnetic field ${\bf B}_{eff}$ well away from 
the edge is parallel to the real magnetic field in Fig.\ref{f1}(a) 
and anti-parallel in Fig.\ref{f1}(b). The apex of each ``fan" 
of energy levels occurs where ${\bf B}_{eff} =0$ for some even 
integer $m$; there according to equations (\ref{Bg}) 
and (\ref{Beff}) 
$E_{m,r} = W$. The different types of composite fermion Landau 
levels are labeled I, II and III. Type I Landau levels are 
silent 
edge modes in the sense that the average 
currents that they carry are 
independent of the electrochemical potential at the edge, 
provided that quasi-equilibrium conditions prevail 
there.\cite{edge} However, Landau levels of types II and III 
carry non-zero average net currents when the 
difference between the composite fermion effective 
electrochemical potentials at opposite edges of the sample
is not zero.\cite{edge}

\section{Current Fluctuations and Fractional Charge}\label{Fluct}

Consider a 2DEG of length $L$ in the {\em x-y} plane connecting
source and drain contacts as depicted in Fig.\ref{f2}. Suppose 
that the magnetic field
${\bf B}$ points in the $z$-direction and 
define an effective vector potential 
${\bf A}_{eff} = (0,\int^x{B_{eff}(u) du},0)$ such that the
effective magnetic field experienced by composite fermions is
given by
${\bf B}_{eff} = \nabla \times {{\bf A}_{eff}}$. The effective
composite fermion Schr\"{o}dinger equation for the $r^{th}$
composite fermion Landau level is then \cite{mass}
\begin{equation}
(\frac{1}{2m^*} (\frac{\hbar}{i} \nabla +e{\bf A}_{eff})^2
+W(x))\psi_{r,k} (x,y) = \epsilon_{r,k} \psi_{r,k} (x,y)
\label{Sch}
\end{equation}
where 
\begin{equation}
\psi_{r,k} (x,y)= e^{iky} X_{r,k} (x).
\label{psi}
\end{equation}
The net electric current in the $y$-direction is then
\begin{equation}
I= - \frac 1 L \sum_{r,k} e {N_{r,k}} v_{r,k}
\label{cur}
\end{equation}
where $N_{r,k}$ and $v_{r,k}$ are the occupation and velocity
of the single particle composite fermion state 
$|r,k>$ and $L$ is the length of the sample.
Since composite fermions obey Fermi statistics, $N_{r,k}$ takes
the values 0 and 1. Fluctuations in the values that the
$N_{r,k}$ take give rise to current noise. However an
important complication 
is that $\epsilon_{r,k}$ in Equation (\ref{Sch})
(and therefore $v_{r,k}$ in Equation (\ref{cur}))
depends on the
specific configuration of occupation numbers $N_{r,k}$ as
will be discussed further below.

In the spirit of mean field theory let us now approximate 
the average electric current $<I>$ by setting every occupation
number ${N_{r,k}}$ in Equation (\ref{cur}) formally 
equal to its 
ensemble average value $<{N_{r,k}}>$ and replacing 
$v_{r,k}$ by the corresponding mean field
velocity function $v_{r,k}^{av}=\frac {1} {\hbar} 
\frac {d\epsilon_{r,k}^{av}} {dk}$, where $\epsilon_{r,k}^{av}$ is
the solution of Equation (\ref{Sch}) with ${\bf A}_{eff}$ and
$W$ calculated using the average occupation numbers $<{N_{r,k}}>$.   
Then
\begin{equation}
<I> = - \frac 1 L \sum_{r,k} e <N_{r,k}> v_{r,k}^{av} 
\label{curav}
\end{equation}
or equivalently
\begin{equation}
<I> = - \frac e h \sum_{r} \int <N_{r,k}> d\epsilon_{r,k}^{av} 
\label{curavint}
\end{equation}
Now let us alter the occupation number $N_{R,K}$ 
of just {\em one} single-particle state $|R,K>$
so that it
differs from the mean $<N_{R,K}>$ that figures 
in Equations (\ref {curav}) and (\ref {curavint}),
and examine the corresponding deviation 
$\delta I_{R,K}$ 
of the current $I$ from $<I>$. (General current fluctuations
will be treated at the end of this Section.)

Bearing in mind 
Equation (\ref{curavint}), $\delta I_{R,K}$ can be written
as 
\begin{equation}
\delta I_{R,K} = \delta I_{R,K}^0  
- \frac e h \sum_{r} \int <N_{r,k}> dU_{r,k}.
\label{curdev}
\end{equation}
Here
\begin{equation} 
U_{r,k}=\epsilon_{r,k}^* -\epsilon_{r,k}^{av} 
\label{U}
\end{equation}
and
\begin{equation}
\delta I_{R,K}^0 = - \frac e L ({N_{R,K}}
-<N_{R,K}>) v_{R,K}^*
\label{bare}
\end{equation}
where $\epsilon_{r,k}^*$ and $v_{r,k}^*$ are the 
self-consistent single particle 
energies and velocities
for the the composite fermion 
system with only $N_{R,K}$ differing from its mean.

Notice that the second term on the RHS of 
Equation (\ref{curdev}) includes the effect of changing
the occupation of the single particle composite
fermion state in Landau Level R with wave vector K
on the currents carried by the {\em other} single 
particle composite
fermion states. It will be seen below that this effect
is very important.

To proceed further analytically I 
will assume the edge currents are of the quasi-equilibrium 
type and that the
edge potential $W(x)$ is so slowly varying with $x$ that 
$<N_{r,k}>$ varies little over the range of values of $k$ 
that contribute appreciably\cite{range} to the integral in Equation 
(\ref{curdev}). 
In Equation (\ref{curdev}), $\int <N_{r,k}> dU_{r,k}$
then becomes $<N_{r,K}> \int dU_{r,k}$.
  
The occupation number fluctuation
$N_{R,K}-<N_{R,K}>$ affects $U_{r,k}$ through the
associated fluctuations in the fictitious electric field
${\bf E}_g$, the Coulomb potential and the effective
magnetic field. Only the first of these contributes to
$\int dU_{r,k}$ because the others are local 
effects and do not affect the integral. 
Thus using 
Equation (\ref{Eg}) for ${\bf E}_g$ yields 
\begin{equation}
\int dU_{r,k}=\Delta U
=\int e\,\delta {\bf E}_g.d{\bf x}
= \pm \frac {mh} e \delta I_{R,K}
\label{intdU}
\end{equation}
where the upper (lower) sign applies when 
${\bf B}_{eff}$ is parallel (antiparallel) to ${\bf B}$.
Notice that the {\em total} current fluctuation
$\delta I_{R,K}$ is used for the integral of the
fluctuation current density in the last term in 
Equation (\ref {intdU}). This ensures that the final
result (\ref {dresscur}) of 
this calculation will include the interactions
between composite fermions that are mediated by the 
fluctuation $\delta {\bf E}_g$ of the fictitious 
electric field  
{\em self-consistently}.

Replacing $\int <N_{r,k}> dU_{r,k}$ by
$<N_{r,K}> \int dU_{r,k}$ in Equation 
(\ref{curdev}) and using (\ref{intdU}) 
yields
\begin{equation}
\delta I_{R,K} = \delta I_{R,K}^0  
\mp m\, \delta{I_{R,K}} \sum_{r} <N_{r,K}>
\label{intermed}
\end{equation}
or
\begin{equation}
\delta I_{R,K} = \frac{\delta I_{R,K}^0}  
{1 \pm m \sum_{r} <N_{r,K}>}
\label{dresscur}
\end{equation}

The physical meaning of this result is that a 
single-particle composite
fermion current fluctuation 
$\delta I_{R,K}^0$ generates a fluctuation in the 
the fictitious electric field that extends over a
distance of a few magnetic lengths. As a 
result a ``kink'' develops in the energy dispersions of
{\em all} of the composite fermion Landau levels
in that vicinity modifying the velocities of the
surrounding composite fermion states. Thus the single
particle current fluctuation is dressed by the 
disturbance
that it generates in the surrounding composite fermion
medium. The 
dressing must be calculated self-consistently since
the current fluctuation induced in the surrounding
medium generates its own contribution to the fluctuation 
in the fictitious electric field which in turn 
affects the current. The self-consistent
calculation
yields the dressed current fluctuation $\delta I_{R,K}$
given by Equation (\ref {dresscur}).

Thus the current fluctuation can be understood as
a dressed quasi-particle associated with the composite
fermion in Landau level R with wave vector K.
The quasi-particle propagates with same
velocity $v_{R,K}^*$ as the composite fermion. 
It is therefore intuitively
appealing to rewrite equation Equation (\ref {dresscur}) 
for the dressed current fluctuation in
a form analogous to Equation (\ref{bare}) for the
``bare" current fluctuation but with
a renormalized quasi-particle charge, i.e.,
\begin{equation}
\delta I_{R,K} = - \frac {e_{R,K}^*} L ({N_{R,K}}
-<N_{R,K}>) v_{R,K}^*
\label{renorm}
\end{equation}
where the dressed quasi-particle charge is given by
\begin{equation}
-e_{R,K}^* = \frac{-e}  
{1 \pm m \sum_{r} <N_{r,K}>}
\label{e*}
\end{equation}

Where the 2DEG is incompressible, $<N_{r,K}>$
is either 0 or 1 depending on whether the $r^{th}$ composite
fermion Landau level is full or empty, and therefore
\begin{equation}
-e^* = \frac{-e}  
{1 \pm m n}
\label{e*bulk}
\end{equation}
where $n=\sum_{r} <N_{r,K}>$ is an integer and $m$ is an even
integer. This agrees with the fractional quasi-particle 
charges that have been derived previously by other methods
for incompressible states
at Jain's filling fractions $\nu=\frac{n}  
{m n \pm 1}$ of a Landau level.\cite{Laughlin}$^,$
\cite{Haldane}$^,$\cite{Halperinfrac}$^,$\cite{GoJa} 
For example, in the bulk for the fractional
filling $\nu=1/3$ of the lowest Landau level, $m=2$, $n=1$ 
and the dressed
quasiparticle charge $-e^*$ given by 
Equation (\ref {e*bulk}) equals $-e/3$. 

On the other hand as the depletion region at the edge
of the sample is approached the filling of the 
composite fermion Landau levels approaches zero. Thus
$\sum_{r} <N_{r,K}> \, \rightarrow 0$ and 
Equation (\ref {e*}) predicts that $-e^*$ tends 
to $-e$.

In the compressible regions between these two extremes
the behavior of the quasi-particle charge can be obtained
from Equation (\ref {e*}) if the behavior of $<N_{r,k}>$ 
is known. For the usual case of quasi-equilibrium edge 
currents this is given by the Fermi function 
\begin{equation}
<N_{r,k}> = \frac 1 {e^{\beta(\epsilon_{r,k}^
{av}-\mu^*_i)}+1}
\label{av}
\end{equation}
where $\mu^*_i$ is the composite fermion electrochemical
potential for the $i^{th}$ (closest) 
edge, $\beta = 1/k_B T$, $T$ is the
temperature and $k_B$ is Boltzmann's constant. 
Johnson-Nyquist noise arises from the compressible regions
in this regime and will be evaluated using the above 
results in Section \ref{Noise}.
 
Since the modification of the velocities of the composite
fermions from $v_{r,k}^{av}$ to $v_{r,k}^*$ by the fictitious
electric field associated with the current fluctuation is
responsible for the dressing of the composite fermion charge
and since the dressed velocity $v_{R,K}^*$ 
appears explicitly in Equation 
(\ref{renorm}) along with $e^*$ it is of 
interest to consider
how large the effect of the fluctuation of the fictitious 
electric field on $v_{R,K}^*$ may be. This may be estimated
as $\sim \frac{\Delta U} {\hbar\, \Delta k}$ where $\Delta U$
is as in Equation (\ref {intdU}) and $\Delta k$ is the
difference in wave vector between states separated by
the magnetic length $a=\hbar/(e B_{eff})$ in real space. 
The result is found to be proportional to $a/L$ and 
is therefore very small for macroscopic samples. The
dressing of the quasi-particle charge on the other
hand is due to the modification by the fictitious
electric field of the velocities of
a {\em large} number (of the order of $L/a$)
of single-particle 
composite fermion states
 and for that reason 
is a very significant 
effect even though the modification of
the individual composite fermion velocities is very
small. 

Finally, it is important to consider whether the above
results which have been obtained by considering the
dressing of a {\em single} single-particle 
current fluctuation are applicable to thermal current 
fluctuations
which involve large numbers of such single-particle
``excitations" simultaneously. This will be done in
two steps:

Firstly, if a current fluctuation is due to any number of
single particle excitations that are widely separated
from each other (by much more than a magnetic length)
in real space then the dressing of each single particle
current fluctuation will occur separately 
from the others and
it follows that Equation 
(\ref {renorm}) becomes
\begin{equation}
\delta I = -\sum_{R,K} \frac {e_{R,K}^*} L ({N_{R,K}}
-<N_{R,K}>) v_{R,K}^*
\label{renormgen}
\end{equation}
where $\delta I$ is the total dressed current fluctuation,
the sum over $R$ and $K$ is over the individual 
single-particle excitations that contribute to the 
current fluctuation and the dressed composite
fermion quasiparticle 
charge $-e_{R,K}^*$ 
is given by Equation (\ref{e*}).

Secondly, if a current fluctuation is due to any number of
single particle excitations that are all so close to
each other in real space that $<N_{r,k}>$ does not
vary appreciably over the entire region containing these
excitations for any 
composite fermion Landau level $r$ then the entire
argument leading to Equation (\ref {dresscur}) still
holds if ${\delta I_{R,K}^0}$ is replaced with
the sum of the ``bare" single particle current 
fluctuations
$-\frac {1} L \sum_{R,K} e ({N_{R,K}}
-<N_{R,K}>) v_{R,K}^*$ 
and ${\delta I_{R,K}}$ is replaced with the total
dressed current fluctuation ${\delta I}$. The result
of this calculation turns out to be also given by
Equations (\ref {renormgen}) and (\ref{e*}).

Since Equations (\ref {renormgen}) and (\ref{e*})
are valid in the two opposite limits of widely 
separated and closely spaced single-particle excitations
discussed above it is reasonable to suppose that they
are valid (or at least a good approximation) in
general, at the level of mean field theory and provided
that the edge potential $W(x)$ and $<N_{r,k}>$ are 
sufficiently slowly varying functions of $x$ and $k$
as has been assumed in the above analysis. 
 
\section{Johnson-Nyquist noise}\label{Noise}

I will now apply the theory developed above 
to the simplest 
relevant experimentally observable 
phenomenon, the Johnson-Nyquist noise on fractional
quantum Hall plateaus where there is no backscattering
of the composite fermions.

Johnson-Nyquist noise is the electric current noise
that occurs in the absence of an applied voltage 
across a conductor connecting two
reservoirs and is due to thermal fluctuations
in the number of charge carriers that pass
through the
conductor per unit time. As depicted schematically
in Fig. \ref {f2}, the reservoirs are assumed
to be effectively short circuited via two large 
capacitances to eliminate any voltage 
fluctuations between them at the frequencies
of interest. 

Current noise is described in terms of the spectral 
density of current-current fluctuations that is 
defined as follows.\cite{LaLi} 
Let the Fourier transform of the current 
fluctuations be 
\begin{equation}
\delta I_\omega = \int_{-\infty}^{\infty} \delta I(t)
\, e^{i\omega t} dt.
\label{FT}
\end{equation}
The current-current correlation function 
$<\delta I(t)\, \delta I(t^\prime)>$ 
depends only on $t-t^\prime$ if
$<\delta I_\omega \, \delta I_{\omega^\prime>}$
is of the form
\begin{equation}
<\delta I_\omega \, \delta I_{\omega^\prime}>
=2\pi F_\omega\, \delta(\omega + \omega^\prime)
\label{delta}
\end{equation}
It then follows that $<(\delta I)^2> \equiv
<\delta I(t)\, \delta I(t)>$ is given by
\begin{eqnarray}
<(\delta I)^2> =
\frac 1 {\pi}
\int_{0}^{\infty} 
F_\omega d\omega
=2 \int_{0}^{\infty} 
F_{2 \pi f} df
\label{corr3}
\end{eqnarray}
where $\omega=2 \pi f$. Thus the the spectral 
density of current fluctuations at 
frequency $f$ is
\begin{equation}
S(f)=2F_{2 \pi f}
\label{S}
\end{equation}
where $F$ is defined by Equation (\ref {delta}).

Let us now rewrite Equation (\ref {renormgen}) as
\begin{equation}
\delta I(t) = \sum_{R,K} \delta I_{R,K}(t)
\label{timedep}
\end{equation}
where $\delta I_{R,K}(t)$ is the dressed 
contribution of
the state with wave vector $K$ in composite 
fermion Landau Level $R$ at time $t$ to the
current fluctuation. The time dependence arises
because the occupation number $N_{R,K}$ fluctuates
taking values 0 and 1 at different times as 
composite fermion wave packets in composite 
fermion Landau level $R$ and centered
on wave vector $K$ are injected 
into the 2DEG
and absorbed from it by the reservoirs. 
In the spirit of
Landauer and Martin's treatment of systems of
ordinary fermions,\cite{Land}$^,$\cite{LM}
I will assume that these
wave packets (which may be occupied or empty) 
are mutually orthogonal and enter
the 2DEG in a regular sequence at times 
$t_{R,K,l}$ spaced by the time
$\delta t_{R,K}=t_{R,K,l+1}-t_{R,K,l}$
that it takes a wave packet to traverse a 
distance $D$ equal to its length. For
the low frequencies $\omega$ of interest 
($\omega \ll 1/\delta t_{R,K}$), 
Equation (\ref {FT}) thus becomes
\begin{equation}
\delta I_\omega = \sum_{R,K}
\int_{-\infty}^{\infty} \delta I_{R,K}(t)
\, e^{i\omega t} dt
=\sum_{R,K,l}\delta t_{R,K} \delta I_{R,K,l} 
e^{i\omega t_{R,K,l}} 
\label{FT'}
\end{equation}
where 
\begin{equation}
\delta I_{R,K,l} = \frac 1 {\delta t_{R,K}}
\int_{t_{R,K,l}}^{t_{R,K,l+1}} \delta I_{R,K}(t)\,dt.
\label{x}
\end{equation}
Assuming that the current fluctuations for different
$R$, $K$ and $l$ are uncorrelated, Equation (\ref{FT'}) 
yields
\begin{equation}
<\delta I_\omega \, \delta I_{\omega^\prime}> 
=\sum_{R,K,l}
 <\delta I_{R,K,l}^2> 
\delta t_{R,K}^2
e^{i(\omega + \omega^{\prime}) t_{R,K,l}}
\label{xxx}
\end{equation} 
Noting that $<\delta I_{R,K,l}^2>$
is independent of $l$ and that for low frequencies 
$\omega, \omega^{\prime} \ll 1/\delta t_{R,K}$,
we may replace $\sum_{l}
e^{i(\omega + \omega^{\prime}) t_{R,K,l}}$ by
$2\pi \delta(\omega + \omega^{\prime})/\delta t_{R,K}$,
Equation (\ref {xxx}) becomes
\begin{equation}
<\delta I_\omega \, \delta I_{\omega^\prime}> 
=2\pi \sum_{R,K}
 <\delta I_{R,K}^2> 
\delta t_{R,K}\,
\delta(\omega + \omega^{\prime})
\label{xxxx}
\end{equation}
which by comparison with Equations (\ref {delta}) and
(\ref {S}) yields
\begin{equation}
S(f)=2 \sum_{R,K}
 <\delta I_{R,K}^2> 
\delta t_{R,K}
\label{spectrum}
\end{equation}
for the spectral density of the current noise. 

In Equation (\ref{spectrum}) $\delta I_{R,K}$ 
is as in Equations (\ref {renorm}) and (\ref {renormgen}) 
but with $L$ replaced by $D$ since the wave packet 
basis is now being used.
Setting $<\!N_{R,K}^2\!> \,
= <\!N_{R,K}\!>$ since $N_{R,K}= $ 0 
or 1, using $|v_{R,K}^*| = D/\delta t_{R,K}$ and 
approximating $v_{R,K}^*$ by $v_{R,K}^{av}$ 
(which is reasonable according to the discussion
in Section \ref{Fluct}),    
Equation (\ref {spectrum}) yields

\begin{equation}
S(f)=2 \sum_{R,K}
 \frac {{e_{R,K}^*}^2} D <\!N_{R,K}\!>
(1-<\!N_{R,K}\!>)\,|v_{R,K}^{av}|
\label{spectrum1}
\end{equation}
Note that since $0 \le \, <\!N_{R,K}\!> \, \le 1 $
the summand in Equation (\ref {spectrum1}) is 
non-negative and is not zero only for states
in compressible regions.
Transforming the sum over $K$ in 
to an energy integral 
yields
\begin{equation}
S(f)=\frac 2 h \sum_{R} \int
 {{e_{R,K}^*}^2} <\!N_{R,K}\!>
(1-<\!N_{R,K}\!>) d\epsilon_{R,K}^{av}
\label{spectrum2}
\end{equation}
Assuming that $<N_{R,K}>$ is of the form given
by Equation (\ref {av}) yields
\begin{equation}
\frac {\partial <\!N_{R,K}\!>} 
{\partial \epsilon_{R,K}^{av}}
= - \beta <\!N_{R,K}\!>(1-<\!N_{R,K}\!>)
\end{equation}
and Equation (\ref {spectrum2}) becomes
\begin{equation}
S(f)=\frac 2 {h\beta} \sum_{R} \int
 {{e_{R,K}^*}^2} d\!<\!N_{R,K}\!>.
\label{spectrum3}
\end{equation}
Finally, noting the form (\ref {e*}) of the
dressed composite fermion charge, Equation
(\ref {spectrum3}) reduces to
\begin{equation}
S(f)=\frac {2e^2} {h\beta} \int
 \frac {dN_{\text{tot}}}{(1 \pm m 
N_{\text{tot}})^2}
\label{spectrum4}
\end{equation}
where $N_{\text{tot}}$ stands for 
$\sum_{r} <\!N_{r,K}\!>$, the local total 
filling of the composite fermion Landau levels.
In keeping with the non-negative character
of the summand in Equation (\ref {spectrum1}),
the convention has been adopted that the 
direction of integration in Equations
(\ref{spectrum2}), (\ref{spectrum3}) and 
(\ref{spectrum4}) is such that 
$d\epsilon_{R,K}^{av}$ , $d\!<\!N_{R,K}\!>$
and $dN_{\text{tot}}$ are always positive, respectively.

Equation 
(\ref{spectrum4}) gives the low frequency 
Johnson-Nyquist noise on fractional 
quantum Hall plateaus where there is no
back scattering of composite fermions.
In general the integral consists of separate
contributions from each region where 
${\bf B}_{eff}$ is parallel or antiparallel to
${\bf B}$ at each of the edges of the 2DEG
that run between the source and drain
in Fig.\ref {f2}.

Let us consider now the case illustrated in 
Fig.\ref {f1}(a) where ${\bf B}_{eff}$ is parallel 
to ${\bf B}$ far from the edges where
the Landau level filling 
$\nu={n_b}/({m_b n_b + 1})$ is a Jain
fraction 
with ${n_b}$ a positive integer and  ${m_b}$ an 
even 
positive integer.

Suppose
initially for simplicity that that only 
the Type III composite fermion Landau levels are
present and that they empty one by one as the edge is 
approached. Then $N_{\text{tot}}$ in Equation
(\ref{spectrum4}) ranges from $0$ at the edge
to $n_b$, the number of occupied 
composite fermion
levels in the bulk, and
the Johnson-Nyquist noise (including the contributions
from both edges of the 2DEG)
is given by
\begin{equation}
S(f)=\frac {4e^2} {h\beta} \int_0^{n_b}
 \frac {dN_{\text{tot}}}{(1 + m_b N_{\text{tot}})^2}
=\frac {4} {\beta} \frac {e^2} {h} 
\frac {n_b} {(m_b {n_b} +1)}
\label{spectrumomlyIII}
\end{equation}
This is just the Nyquist formula $S(f)=4{k_B}T G$ since
on a quantum Hall plateau the two-terminal conductance
$G$ is equal to the quantized Hall conductance 
$G_H = \frac {e^2} {h} 
\frac {n_b} {(m_b {n_b} +1)}$.

Now let us consider the contribution of the ``silent''
\cite{edge} Type I composite fermion 
Landau levels in Fig.\ref {f1}(a) to $S(f)$. For
these $m=m_b +2$. Let us suppose that the
crossover from the Type III to the Type I states
occurs at a local Landau level filling 
fraction $\nu_c$ where $N_{\text{tot}}={n_c}$ 
for the Type III states and at 
$N_{\text{tot}}={n_c^\prime}$ 
for the Type I states. Then
\begin{equation}
\frac {n_c}{m_b n_c + 1}=
\nu_c =\frac {n_c^\prime}{(m_b+2) n_c^\prime - 1}
\label{cross}
\end{equation}
and
\begin{equation}
S(f)=\frac {4e^2} {h\beta} (\int_{n_c}^{n_b}
 \frac {dN_{\text{tot}}}{(1 + m_b N_{\text{tot}})^2}
+\int_{n_c^\prime}^{\infty}
 \frac {dN_{\text{tot}}}{(1 - (m_b +2) N_{\text{tot}})^2}
+\int_{0}^{\infty}
 \frac {dN_{\text{tot}}}{(1 + (m_b +2) N_{\text{tot}})^2})
\label{spectrumIII+I}
\end{equation}
where the first integral is the contribution of the
Type III states, the second (third) integral
is that of the Type I states
for which ${\bf B}_{eff}$ is antiparallel (parallel) to
${\bf B}$. The limits of integration where $N_{\text{tot}}
=\infty$ correspond to ${\bf B}_{eff}$ passing
through zero. Evaluation of (\ref {spectrumIII+I}) using
the relation (\ref {cross}) between ${n_c}$ and 
${n_c^\prime}$ once again yields the same 
Nyquist expression
\begin{equation}
S(f)
=\frac {4} {\beta} \frac {e^2} {h} 
\frac {n_b} {(m_b {n_b} +1)}
\label{spectrumomlyIIIagain}
\end{equation}
as was obtained above by considering the simpler model with
{\em only} Type III composite fermion Landau levels. Thus
although the silent Type I levels contribute to 
the Johnson- Nyquist noise, remarkably, 
the {\em total} of the 
Type I and Type III Johnson-Nyquist
noise is the same whether the   
Type I levels are present in the model or not.

The Johnson-Nyquist noise on quantum Hall plateaus
for the case shown in 
Fig.\ref {f1}(b) where ${\bf B}_{eff}$ is antiparallel 
to ${\bf B}$ far from the edges can be calculated
similarly.
Here
the Landau level filling in the bulk 
is the Jain fraction
$\nu={n_b}/({m_b n_b - 1})$ 
with ${n_b}$ a positive integer and  ${m_b}$ an 
even positive integer. In this case {\em only} the
Type I levels contribute to the Johnson-
Nyquist noise and the result of evaluating
Equation 
(\ref{spectrum4}) is
\begin{equation}
S(f)
=\frac {4} {\beta} \frac {e^2} {h} 
\frac {n_b} {(m_b {n_b} -1)}
\label{spectrumBeffneg}
\end{equation}
once again in agreement with the Nyquist formula
$S(f)=4{k_B}T G$.
The result for $S(f)$ is once again the same 
whether only 
the states with $m=m_b$ or both the states with
$m=m_b$ and $m=m_b +2$ are included in the model.

\section{Conclusions}\label{Conc}

In this paper it has been shown that interactions between
composite fermions must be included in  
composite fermion theories of electric current noise.
 
The important interactions were found 
to be those mediated by the 
fictitious electric field generated by composite
fermion current fluctuations. Their effects must
be calculated self-consistently. This was done
analytically in mean field theory for systems with 
smooth edge potentials. 

The interactions were found to
renormalize the current fluctuations and the charges of
the quasi-particles associated with composite fermions,
making the quasiparticles fractionally charged. 
Analytic expressions were obtained for the
charges of the quasiparticles in both the 
incompressible and compressible regions of composite
fermion systems. In the incompressible regions the
calculated fractions agree with previous theories. In the
compressible regions the quasiparticle charge varies 
smoothly with position. It tends to zero where the 
effective magnetic field vanishes and becomes 
equal to the electron
charge where the composite fermion density becomes zero.   

This theory of composite fermion current fluctuations 
and quasi-particle charge was then applied to an 
observable phenomenon by calculating
the Johnson-Nyquist noise on fractional quantum Hall
plateaus where there is no back-scattering. The results
obtained were in agreement with the  Nyquist formula
$S=4{k_B}T G$. 

The Nyquist formula is usually
derived from the fluctuation-dissipation theorem
\cite{Imry} so that 
this result is not surprising. However the present 
derivation of the Nyquist formula is of interest because   
it provides a good test of the mean field theory
of current fluctuations and quasiparticle charge in this
{\em interacting} system, and because it provides significant 
insights into the relationship between quasi-particle 
fractional charge, composite fermion edge states and 
current noise. Some of these are:

The Johnson-Nyquist noise in the fractional quantum Hall 
regime arises from {\em compressible} regions
of the 2DEG. It is therefore {\em not} primarily due to 
quasi-particles having the usually quoted values of 
the fractional 
charge that correspond to 
incompressible fractional quantum Hall states. For
example for the $\nu=1/3$ quantum Hall state the
Johnson-Nyquist noise arises from quasiparticles
whose charges are for the most part {\em not} equal to
$1/3$ of the
electron charge but span a wide range of values.

Compressible edge channels associated with 
both the normal Type III and silent Type I composite 
fermion Landau level
edge states\cite{edge} 
contribute to the Johnson-Nyquist noise
when they are present at the edge, and their 
contributions sum to yield the Nyquist formula.  

Shot noise differs from Johnson-Nyquist noise in that
a finite voltage is applied to the sample and a 
non-zero average current flows. {\em Both}
incompressible and compressible regions at the
edge contribute to the current noise in this case. 
A full discussion
of shot noise is beyond the scope of this paper,
however it is already evident from the theory developed
above that quasi-particles with a range of different 
values of the quasi-particle charge associated with
compressible and incompressible
strips at the sample edges will contribute
to shot noise. Thus the present work suggests that 
the arguments that have been
used recently\cite{Saminadayar}$^,$
\cite{de-Picciotto} to infer a {\em single} value
of the fractional quasi-particle charge from shot
noise measurements may need some refinement.

This work was supported by NSERC of Canada.

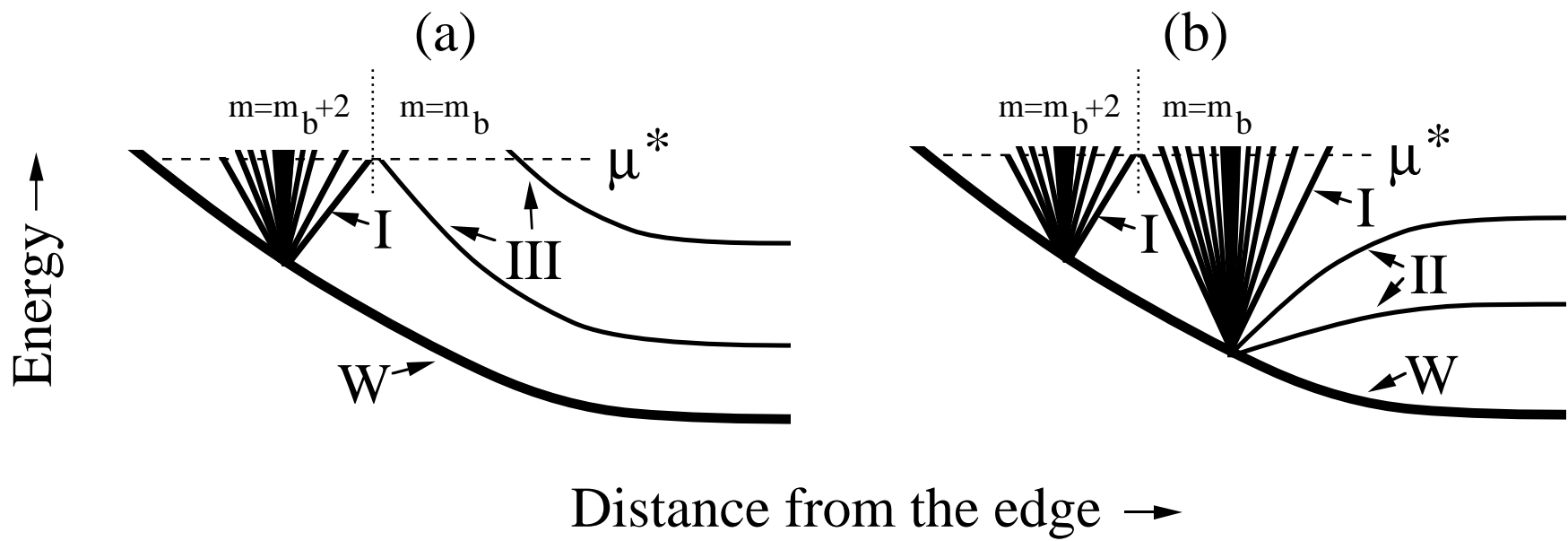
\begin{figure}
\caption{Schematic drawing of the composite fermion Landau level 
structure near an edge. The effective magnetic field 
${\bf B}_{eff}$ well away from the edge is parallel to the 
real magnetic field in (a) and anti-parallel in (b). The apex 
of each ``fan" of energy levels occurs where ${\bf B}_{eff}=0$ 
for an even integer $m$, where $m$ is the number of fictitious 
flux quanta attached to each electron. $m_b$ is the number
of fictitious 
flux quanta attached to each electron in the bulk,
far from the edges of
the sample. Vertical dotted lines delimit the regions with
different $m$.
The different 
types of composite fermion edge states are labeled I, II and III.}
\label{f1}
\end{figure}
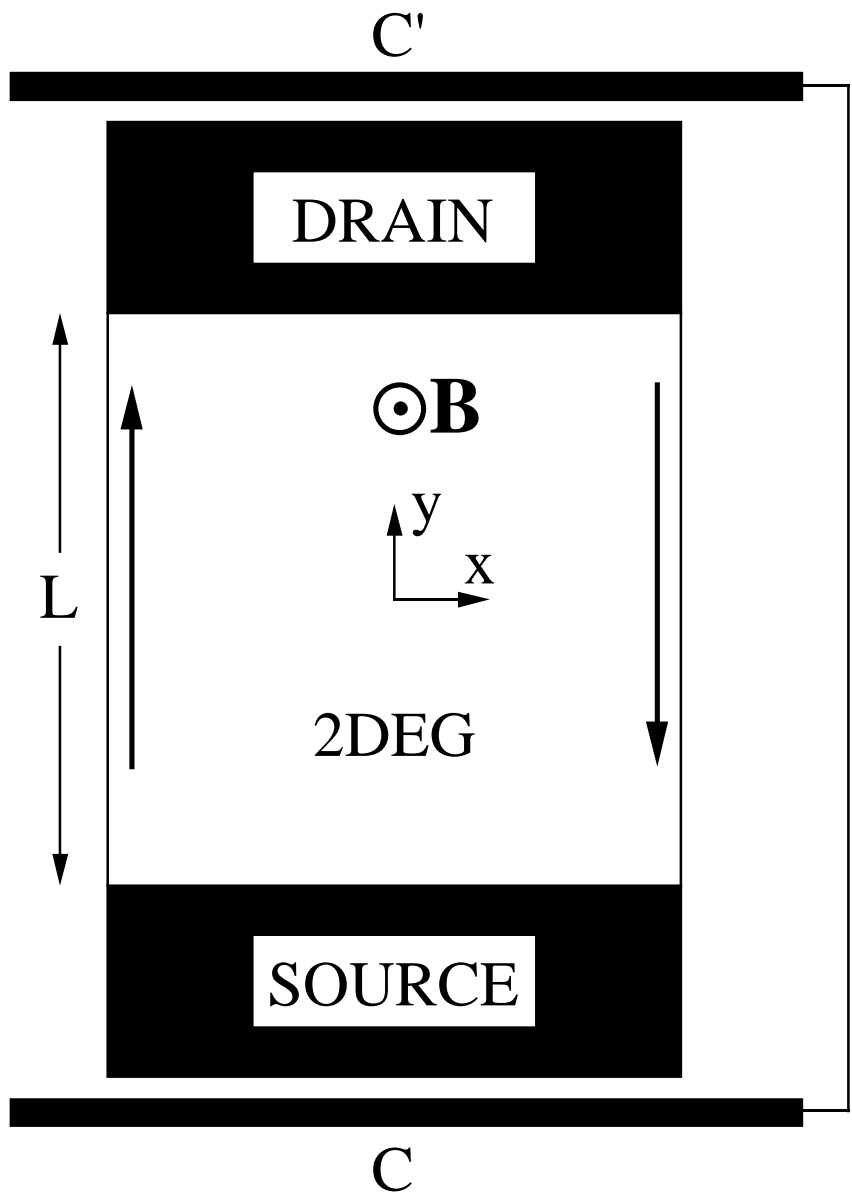
\begin{figure}
\caption{Schematic of a 2DEG of length $L$ connecting source and
drain reservoirs in a magnetic field. The reservoirs are 
effectively short circuited by the capacitors C and C' at the
frequencies of interest. Arrows indicate the direction of
propagation of edge states.}
\label{f2}
\end{figure}

\begin{references}
\bibitem{range}The integrand is appreciable over a
range of states $|r,k>$ that are within 
a few magnetic lengths
of the state $|R,K>$ in real space. $L$ is considered to
be large enough that there are many single-particle states
in this range.
\end{references}
\end{document}